# NICT's versatile miniaturized lasercom terminals for moving platforms

Alberto Carrasco-Casado, Koichi Shiratama, Phuc V. Trinh, Dimitar Kolev,
Femi Ishola, Tetsuharu Fuse, Hiroyuki Tsuji, and Morio Toyoshima

*Space Communication Systems Laboratory, Wireless Networks Research Center,*
*National Institute of Information and Communications Technology (NICT), Japan*

*Abstract*—With the goal of meeting the diverse requirements of many different types of platforms, ranging from small drones to big satellites, and being applied in a variety of diverse scenarios, ranging from fixed terrestrial links to moving platforms in general, and operating within a wide range of conditions and distances, the Japanese National Institute of Information and Communications Technology (NICT) is currently working towards the development of a series of versatile miniaturized free-space laser-communication terminals. By choosing the appropriate terminal configuration for any given scenario, the basic conditions of operations can be satisfied without the need of customization, and the adaptive design of the terminals can close the gap to achieve an optimum solution that meets the communication requirements. This paper presents NICT`s current efforts regarding the development of this series of lasercom terminals and introduces the first prototypes developed for validation and test purposes.

## I. INTRODUCTION

Free-space optical communications have the potential to bring the bandwidth of optical fibers to moving platforms, greatly enhancing their communication capabilities. This communication technology has already become mature after numerous demonstrations during the last few years [1]. The Japanese National Institute of Information and Communications Technology (NICT) has been one of the pioneers, leading some of the most significant demonstrations over the last three decades. In 1994, NICT carried out the first space-to-ground downlinks using the GEO satellite ETS-VI [2]. Ten years later, NICT participated in the first LEO-to-ground demonstration with the JAXA's LEO satellite OICETS in 2006 [3]. Another ten years later, NICT mounted the first lasercom terminal onboard a microsatellite to perform LEO-ground laser communications and quantum key distribution (QKD) experiments using the LEO satellite SOCRATES in 2014 [4]. This manuscript introduces the ongoing efforts in NICT towards the development of a new series of miniaturized laser-communication (lasercom) terminals with the goal of closing the gap between demonstration and operational deployment.

## II. USE CASES FOR NICT'S LASERCOM TERMINALS

Free-space laser communication is expected to play a key role to cope with the demanding bandwidth requirements of 5G (and beyond) networks to support an increasing number of wireless terminals disseminated throughout the world and generating an unprecedented amount of data. For this purpose, practical and versatile lasercom systems will be necessary to be developed and deployed in real scenarios as soon as possible. NICT intends to meet this requirement by developing a series of communication terminals that can fit a variety of platforms and scenarios to fulfill the requirements of different use cases. Fig. 1 shows several of these use cases, where NICT is currently planning to demonstrate and deploy these terminals.

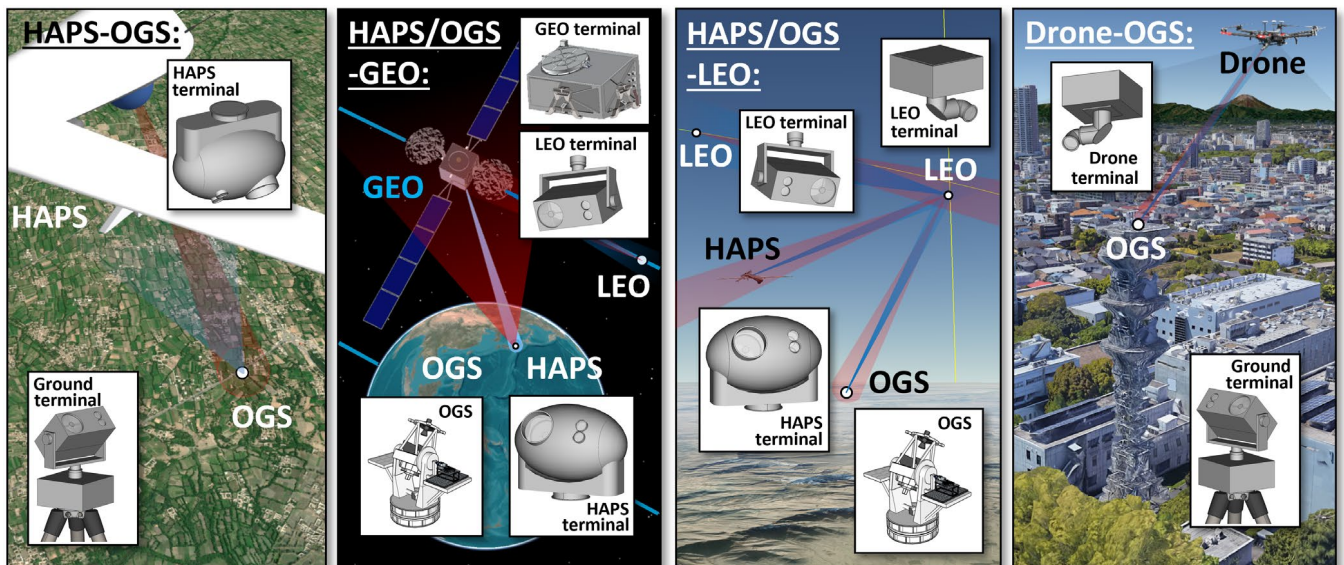

Fig. 1. Use-case examples of NICT's lasercom terminals in a variety of different scenarios and platforms.

Table I. Optical-head assemblies (OHA) of NICT's space laser-communication terminals.

| | **HICALI terminal** | **FX terminal** | **ST terminal** |
|---|---|---|---|
| **Max-range scenario** | GEO-ground (2 ways) | LEO-GEO (1 way), LEO-LEO (2 ways) | LEO-ground (1 way) HAPS-ground (2 ways) |
| **Aperture size** | 15 cm | 9 cm | 3 cm |
| **Mass (OHA)** | ~80 kg | ~8 kg | ~4 kg |
| **Data rate (max)** | 10 Gbps | 10 Gbps (+ future option of 100's Gbps) | |
| **Spectral band** | C-band | | |
| **Direction select.** | Polarization + wavelength | Wavelength (+ polarization compatible) | |
| **Field of regard** | 360° × ±10° | 360° × ±90° | 360° × ±90° |
| **CAD image (OHA)** | 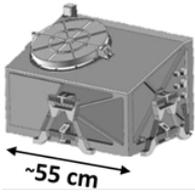 | 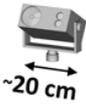 | 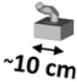 |

The first scenario (fig. 1, left image) includes the use of High-Altitude Platform System (HAPS) to distribute RF communication signals to a large number of mobile users over a wide area of terrain on the ground, taking advantage of the high altitude where HAPS can fly (up to 20 km) during long periods of time (up to several months). These unique characteristics make HAPS very suitable platforms to give continuous coverage to numerous ground users in a more effective way than communication towers and with less demanding requirements than communication satellites. In this use case, laser communications can provide the high-speed feeder links to support the user communications on the ground.

The second scenario (fig. 1, middle-left image) includes the use of GEO (or MEO) satellites as relays between a ground station and ground users or satellite users in LEO. The former use case is currently being implemented by NICT within the ETS-IX project, which will use a 10-Gbit/s bidirectional optical feeder link to provide 100-Mbit/s to user links through Ka band [5]. In the latter use case, the GEO satellite gives service to satellite users, which generally require transmitting large amounts of data to the ground. When high link availability is a critical requirement, using a relay satellite is a useful alternative to waiting to pass over an operational ground station [6]. HAPS or commercial airplanes are another possible user for the GEO relay.

The third scenario (fig. 1, middle-right image) includes the use of LEO satellites, a popular platform which has experienced a considerable growth due to the cost reduction of developing and launching small satellites and CubeSats [7]. There are two basic uses for laser communications from LEO satellites, i.e. direct-to-Earth downlinks or intersatellite links. In general, both implementations have different needs, with the former requiring intermittent and mostly-unidirectional links, and the latter requiring continuous and bidirectional communications since their expected application is LEO constellations.

The fourth scenario includes the use of drones, another platform gaining popularity with big potential in a variety of applications that NICT is currently investigating [8]. For example, single or multiple drones can be used as base stations and operate independently or cooperatively to provide communication services to mobile users while maintaining optical feeder links with the ground. These flexible platforms can be rapidly deployed in any situation such as natural disaster or temporary disruption and they allow continuous or quasi-continuous operation by replacing their batteries or by using tethered systems, and in the future by wireless power transfer.

III. NICT'S SERIES OF MINIATURIZED LASERCOM TERMINALS

Table 1 shows all the lasercom terminals that are currently being developed by NICT. The first one is HICALI [9], which development was already completed, and it is currently under qualification and test, prior to being integrated and embarked in the GEO satellite ETS-IX, to be launched in 2023. HICALI was designed specifically for a very well-defined and fixed scenario, i.e., bidirectional 10-Gbit/s links between ground and GEO orbit. The other two terminals are called Full Transceiver (FX) and Simple Transmitter (ST). Unlike HICALI, these terminals were not designed for a specific scenario or platform, but with the goal to make them as versatile as possible. The maximum distance range defined in the table refers to the worst-case scenarios in which the terminals can work, but they are designed to adapt their operation to support other use cases, e.g., with significantly shorter distances as well.

The FX terminal (fig. 2, right), as its name indicates, was designed to meet the requirements of high-speed bidirectional communications at long distances. Bidirectional communications can be supported up to several thousands of kilometers in a LEO-LEO scenario such as LEO constellations. However, the terminal can also support one-way communications up to GEO to support users such as LEO observation satellites that need to transmit data to a GEO relay with very-high availability. This specific application was the original use case of the CubeSOTA concept, which was conceived to work together with ETS9-HICALI [10]. The key component of the FX terminal, which enables it for high-speed bidirectional long-range communications, is the 9-cm miniaturized telescope that allows collimating a very-narrow beam to cover long distances with small geometrical losses, as well as providing enough receiving gain to close the link at high speed.

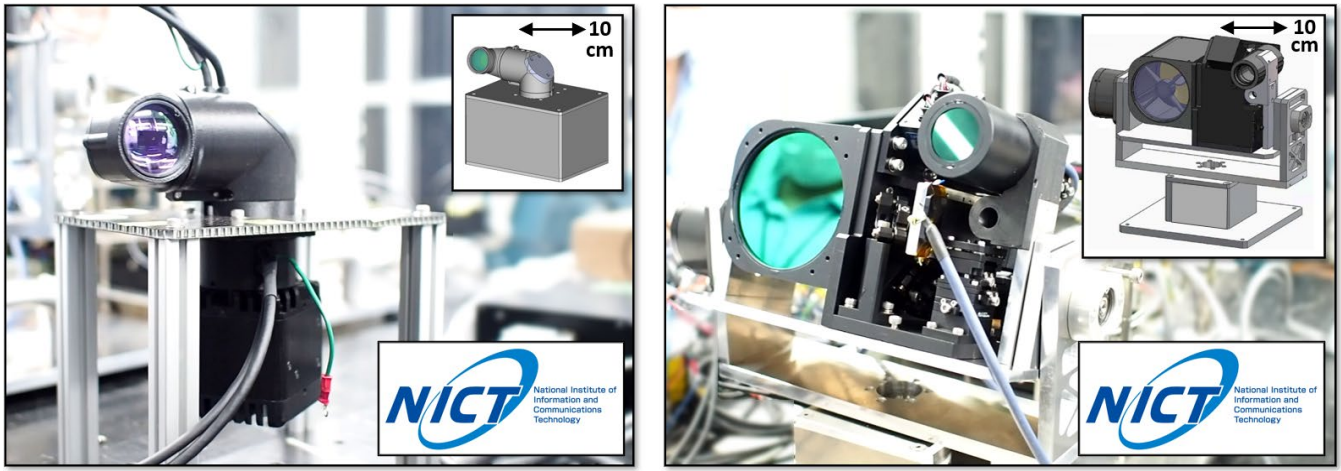

Fig. 2. Early prototypes of ST terminal (left) and FX terminal (right) for preliminary validation and tests.

The ST terminal (fig. 2, left) is a further-miniaturized version of the FX terminal with a compact gimbal design and a smaller aperture, although enough to collimate a laser beam to support LEO-ground downlinks, which was the original concept for this terminal. In the ST terminal, the smaller aperture allows to significantly miniaturize the gimbal at the cost of the smaller receiving gain, which sacrifices the high-speed bidirectional operation at distances such as LEO-ground. However, one-way communication can meet the requirement of many LEO satellites, which needs to downlink data to the ground at high speed, but not so much to receive data in the satellite. The ST terminal can support high-speed bidirectional communications too when the distance is shorter, such as LEO-LEO in dense satellite constellations, or HAPS-ground scenarios, because the internal optical configuration is equivalent to the FX terminal, including the receiving part.

The classical approach when designing a lasercom terminal is reaching an optimum design for a specific demonstration or mission. However, this makes it very difficult to leverage the design for other applications where the conditions may differ. In that case, if the requirements change, the terminal must be redesigned, or at least customized, which is costly. NICT's strategy is to come up with a design that can be reutilized in different circumstances by its internal adaptive operation and a minor configuration of its components. Depending on the scenario and platform, the basic configuration is established by selecting one type or the other, FX or ST. Then, the modem can be selected between a 10-Gbit/s type and a 100's-Gbit/s type, to further refine the requirements of the scenario. When the basic terminal configuration has been determined for any given platform/scenario, the terminal itself can adapt its performance to the varying conditions of the links by internal adaptive techniques.

IV. NICT'S LASERCOM TERMINAL CHARACTERISTICS

Although the FX and ST terminals have a different external appearance, both share the same basic configuration of the internal optics. Both terminals include a fine pointing and tracking system to enable single-mode fiber coupling for the receiver subsystem, as well as a point-ahead mechanism in the transmitter subsystem to enable scenarios involving long distance and narrow beam divergence, where the transmitted and received beams are angularly separated due to the finite speed of light. Both fine-pointing mechanisms have an angular accuracy better than ±1 µrad and the bandwidth of the closed loop is better than 500 Hz (-3 dB). Both terminals have coarse-pointing detectors to close the control loop with the 2-axis gimbals with a full-angle field of view of 1° to allow fast initial acquisition of counter terminals.

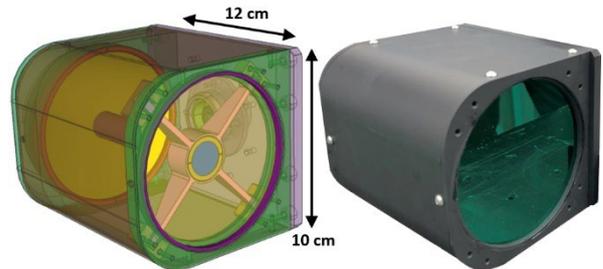

Fig. 3. 3D model and prototype of the 9-cm telescope.

One of the most-critical parts of the FX terminal is the telescope (fig. 3). Since the terminal was designed to be compatible with the CubeSat's form factor, the maximum realistic optical aperture was determined to be 9 cm. The developed telescope is Cassegrain type with 40× magnification and a flat mirror hidden behind the secondary one to steer the beams towards/from the lateral side, where the internal optics are located. The design and development of the telescope was already completed, and partial space qualification has been carried out, including analysis of the wavefront error using a Fizeau interferometer (fig. 4, middle) for different stages of thermal cycling (fig. 4, middle) and vibration (fig. 4, left), getting an error smaller than λ/19 RMS at 1550 nm. The full telescope group, including primary mirror, secondary mirror, folded mirror, and collimator lens, has a total transmission of 93%.

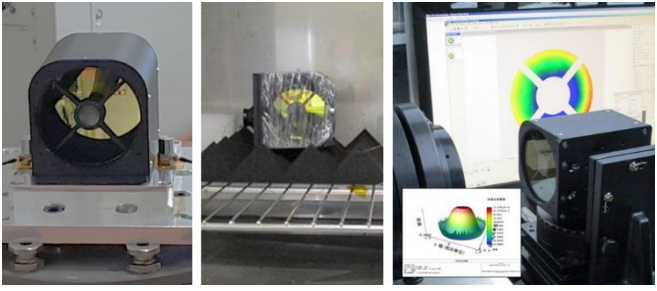

Fig. 4. Telescope's vibration and thermal-cycling tests.

A high-power optical amplifier was designed and developed as well to be integrated with the terminals (fig. 5). For the same reason as in the telescope, i.e. to be compatible with the CubeSat form factor, a maximum optical power of 2 W was chosen. The miniaturized EDFA fits within 1U and has a height of 2.5 cm. The design of this amplifier was based on a LEO-to-ground scenario, where the required operation time is limited to about 10 minutes during each downlink. This condition makes it possible to maintain a transmitted optical power over 2 W during the full pass with such a small device without any additional heat dissipation system. For longer operation times, the same EDFA can operate up to 1 hour over 1.6 W, and other methods are being investigated to extend this operation even further.

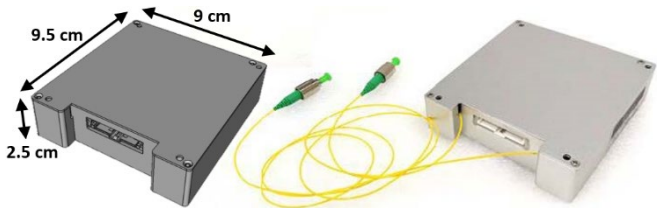

Fig. 5. 3D model and flight model of CubeSat's EDFA.

This amplifier was already fully qualified for space environment, including vibration (fig. 6, left image), shock, thermal vacuum (fig. 6, right image), and total ionization dose tests according to the ISO 19683:2017 and ISO 15864:2004 standards. A low-noise amplifier (LNA) version of the same EDFA will be developed as well with the same format factor to allow bidirectional operation. In this regard, a miniaturized bidirectional 10 Gbit/s modem is also being developed to support 2-way communications, including a high-capacity memory capable of storing the amount of data that a LEO satellite can downlink to a ground station during a typical pass. The modem includes interleaving and FEC to achieve error-free communications under atmospheric turbulence. An upgraded version of an ultra-high-speed modem is currently under development for less-stressed scenarios such as HAPS-ground, with the target to achieve hundreds of Gbit/s with single wavelength and Tbit/s regime with WDM, which the terminals support by design.

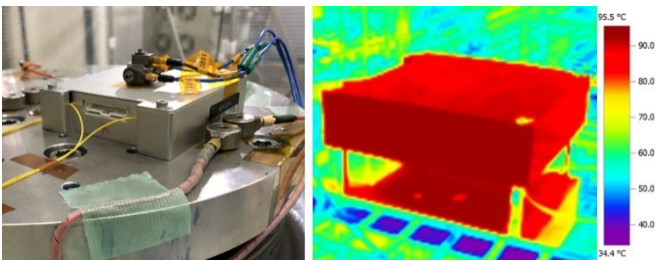

Fig. 6. EDFA's vibration and thermal-cycling tests.

The discrimination between transmission and reception is based on wavelength only, as shown in table I. A sufficiently-wide band gap is kept separating both communication directions to achieve enough optical isolation by spectral filtering without the need to use polarization too. In each band, only one wavelength will be used initially, with the goal of allowing up to 4 WDM channels per direction in the future. Not using polarization for transmit/receive discrimination enables the possibility of doubling the communication throughput by transmitting 2 orthogonal polarizations in each direction when using digital coherent detection techniques, as is the case of the 100's-Gbit/s-class modem. Additionally, polarization-dependent discrimination requires very-precise polarization-angle alignment to avoid extra losses and crosstalk, which adds complexity to the terminals-development process. Nevertheless, the transmitting sides of both terminals allow preparing circularly polarized signals in order to be compatible with polarization-dependent ground stations, as is the case of most NICT's ground stations, since they were designed for the SOTA and HICALI missions [11], which both used orthogonally circular polarizations for transmit/receive discrimination.

Fig. 2 shows the early prototypes of both terminals, which were developed in 2020, for preliminary validation purposes. Upgraded versions of both terminals were designed in 2021 and they are currently being developed with the goal to achieve full functionality and autonomous operation as well as further miniaturization. The final prototypes will be completed in 2022 and will be validated in experiments using a flying drone. The ST terminal will be embarked onboard the drone platform and the FX terminal will play the role of the ground station, separated by several kilometers of distance. Full high-speed bidirectional communications will be tested, including automatic GPS position exchange and discovery, initial counter-terminal acquisition, and fine tracking and pointing during drone's motion. Preliminary tests have already started using a basic test terminal onboard the drone and a basic ground station. Further demonstration plans for the terminals include HAPS version for HAPS-ground and HAPS-HAPS operation, CubeSat's version with no gimbal (based on body pointing), and gimballed version for microsatellite.

## V. Conclusion

This paper introduces the current efforts in NICT towards the development of a new series of miniaturized lasercom terminals with the goal of meeting the requirements of many different scenarios and platforms. The early prototypes of both terminals were shown, some key already-completed subsystems of the terminals were presented, and the basic characteristics of the final terminals were described. Lastly, the basic schedule as well as development and demonstration plans were briefly introduced.

References

[1] A. Carrasco-Casado and R. Mata-Calvo, Space Optical Links for Communication Networks, in: B. Mukherjee, I. Tomkos, M. Tornatore, P. Winzer, Y. Zhao (Eds.), Springer Handbook of Optical Networks, Springer Handbooks, Springer, Cham, 2020, pp. 1057–1103.

[2] K. Araki et al., Performance evaluation of laser communication equipment onboard the ETS-VI satellite, Proc. SPIE Vol. 2699, Free-Space Laser Communication Technologies VIII, San Jose, United States, 1996, 27 January – 2 February.

[3] M. Toyoshima et al.: Results of Kirari optical communication demonstration experiments with NICT optical ground station (KODEN) aiming for future classical and quantum communications in space, Acta Astronautica, 74, 40 – 49 (2012).

[4] A. Carrasco-Casado et al., LEO-to-ground optical communications using SOTA (Small Optical TrAnsponder) – Payload verification results and experiments on space quantum communications, Acta Astronautica, 139, 377 – 384 (2017).

[5] M. Toyoshima, Hybrid High-Throughput Satellite (HTS) Communication Systems using RF and Light-Wave Communications, 2019 IEEE Indian Conference on Antennas and Propogation (InCAP), Ahmedabad, India, 2019, 19 – 22 December.

[6] A. Carrasco-Casado et al., Intersatellite link between CubeSOTA (LEO CubeSat) and ETS9-HICALI (GEO satellite), IEEE International Conference on Space Optical Systems and Applications (ICSOS), Portland, United States, 2019, 14 – 16 October.

[7] A. Carrasco-Casado et al., Optical Communication on CubeSats – Enabling the Next Era in Space Science –, IEEE International Conference on Space Optical Systems and Applications (ICSOS), Okinawa, Japan, 2017, 14 – 16 November.

[8] P. V. Trinh et al., Experimental Channel Statistics of Drone-to-Ground Retro-Reflected FSO Links With Fine-Tracking Systems, IEEE Access, 9, 137148 – 137164 (2021).

[9] Y. Munemasa et al., Critical Design Results of Engineering Test Satellite 9 Communications Mission: for High-Speed Laser Communication, "HICALI" mission, IAC-19-F2.2.3, 70th International Astronautical Congress, Washington D.C., United States, 2019, 21 – 25 October.

[10] A. Carrasco-Casado et al., Development of a miniaturized laser-communication terminal for small satellites, IAC-21-B2.2, 72th International Astronautical Congress, Dubai, United Arab Emirates, 2021, 25 – 29 October.

[11] Dimitar Kolev et al., Preparation of High-Speed Optical Feeder Link Experiments with "HICALI" Payload, 11993-35, SPIE Photonics West, San Francisco, California, United States, 2022, 22 – 27 January.